\documentclass[aps,prl,twocolumn]{revtex4-2}
\usepackage{graphicx,color}
\usepackage{hyperref}
\usepackage{amsmath} 
\usepackage{amsfonts}

\usepackage[normalem]{ulem}
\newcommand{\be}{\begin{equation}} 
\newcommand{\ee}{\end{equation}} 
\newcommand{\bea}{\begin{eqnarray}} 
\newcommand{\eea}{\end{eqnarray}}

\newcommand{\la}{\langle}
\newcommand{\ra}{\rangle}

\begin{document}
\title{Non-reciprocal interactions preserve the  universality class of  Potts model}
\author{ Soumya K. Saha and P. K. Mohanty}
\email{pkmohanty@iiserkol.ac.in}
\affiliation {Department of Physical Sciences, Indian Institute of Science Education and Research Kolkata, Mohanpur, 741246 India.}
\date{\today}

\begin{abstract}
We study the $q$-state Potts model on a square lattice with directed nearest-neighbor spin-spin interactions that are inherently non-reciprocal. Both equilibrium and non-equilibrium dynamics are investigated. Analytically, we demonstrate that non-reciprocal interactions do not alter the critical exponents of the model under equilibrium dynamics. In contrast, numerical simulations with selfish non-equilibrium dynamics reveal distinctive behavior. For $q=2$ (non-reciprocal non-equilibrium Ising model), the critical exponents remain consistent with those of the equilibrium Ising universality class. However, for $q=3$ and $q=4$, the critical exponents vary continuously. Remarkably, a super-universal scaling function—Binder cumulant as a function of $\xi_2/\xi_0$, where $\xi_2$ is the second moment correlation length and $\xi_0$ its maximum value—remains identical to that of the equilibrium $q=3,4$ Potts models. These findings indicate that non-reciprocal Potts models belong to the superuniversality class of their respective equilibrium counterparts.
\end{abstract}
\maketitle
The study of non-reciprocal interactions, where the influence of one component on another is asymmetric, has emerged as a frontier in statistical physics \cite{Ramaswamy2022} due to their widespread occurrence across diverse systems, including active matter \cite{Tailleur2023, Bechinger2016}, condensed matter \cite{Liu2023, Reisenbauer2024, Raskatla2024, Downing2021}, biological systems \cite{Klapp2023, Hickey2023, Bhattacherjee2023}, and social systems \cite{Xie2024}. Unlike conventional reciprocal interactions, non-reciprocal systems exhibit a range of novel phenomena, such as self-assembly \cite{Navas2023}, topological phases \cite{Topfer2022, Rouzaire2024}, hyperuniform states \cite{Chen2024}, non-reciprocal synchronization \cite{Ho2024}, time-crystalline order \cite{Hanai2024}, self-organization \cite{Osat2023}, and unconventional phase transitions \cite{Fruchart2021, Kryuchkov2018, Saha2020, Knezevic2022}.

In stochastic systems, non-reciprocal interactions drive complex dynamics. For instance, asymmetric neural networks studied by Sompolinsky and Kanter \cite{Sompolinsky1986} exhibit altered temporal association dynamics. Similarly, non-reciprocal interactions are observed in colloidal systems with tunable anisotropy \cite{Navas2023} and multicellular assemblies with asymmetric cell-cell communication \cite{Bhattacherjee2023}. Such interactions are also evident in catalytically driven particles \cite{Mandal2024} and soft matter systems \cite{Kryuchkov2018}.

Although non-reciprocal interactions appear to violate Newton’s third law of equal and opposite reaction \cite{Ivlev2015}, they have been experimentally demonstrated in systems such as micro-particles in anisotropic plasmas \cite{Lisin2020}, motile particles in elastic media \cite{Gupta2022}, programmable robots \cite{Fruchart2021}, optically levitated nanoparticles \cite{Rieser2022}, micro-scale oil droplets in surfactant solutions \cite{Meredith2020}, and ions interacting in plant cells \cite{Kronzucker2008}. Often, these interactions emerge as effective forces driven by collective activity, such as the persistent or chiral motion of active particles \cite{Tailleur2023, Knezevic2022} or molecular crowding in chemical systems. Recent studies even suggest that confined water can induce directed non-reciprocal interactions \cite{Cruz2020}.

Despite extensive research in active matter systems \cite{Tailleur2023}, the implications of non-reciprocal interactions for equilibrium and non-equilibrium critical behavior remain poorly understood. In a recent study, Chen et al. \cite{Chen2024} demonstrated emergent chirality and hyperuniformity in active mixtures mediated by non-reciprocal interactions. Similarly, collective flocking dynamics have been observed in active fluids with non-reciprocal orientational interactions \cite{Knezevic2022}. Non-reciprocal interactions can enhance heterogeneity \cite{Carletti2022}, stabilize dynamical phases \cite{Chiacchio2023}, and facilitate particle escape from kinetic traps \cite{Osat2024}.

To elucidate the nature of phase transitions, such as phase separation, crowding, flocking, or condensation in the presence of non-reciprocal interactions, several models have been proposed. The non-reciprocal Cahn-Hilliard model \cite{Saha2020} provides a framework for understanding pattern formation in non-equilibrium systems. Time-delayed interactions, as studied by Durve et al. \cite{Durve2018}, further highlight their significance in active particle condensation. Additionally, non-reciprocal extensions of Ising \cite{Avni2023, Rajeev2024}, XY \cite{Rouzaire2024}, and Heisenberg \cite{Bhatt2023} spin models have revealed modifications to phase behavior and scaling exponents.

In this Letter, we analyze the equilibrium and non-equilibrium dynamics of $q$-state Potts models with directed, non-reciprocal nearest-neighbor interactions on a square lattice. Our analytical results demonstrate that non-reciprocal interactions preserve critical exponents in equilibrium, maintaining universality class invariance. Numerical simulations with selfish non-equilibrium dynamics show that for $q=2$ (Ising-spins), the critical exponents align with those of the equilibrium Ising model. For $q=3$ and $q=4$, continuously varying critical exponents are observed. However, the Binder cumulant reveals  that it is  super-universal scaling function of $\xi_2/\xi_0$, which matches the equilibrium Potts model. These findings suggest the existence of a superuniversality class \cite{ Delfino2017,Bonati_PRL2019, Indranil_2023, Kanti_2024, Aikya2024} bridging the critical behaviour of 
 non-reciprocal Potts model with  equilibrium and non-equilibrium dynamics.

\begin{figure}[t]
    \centering
    \includegraphics[width=0.8\linewidth]{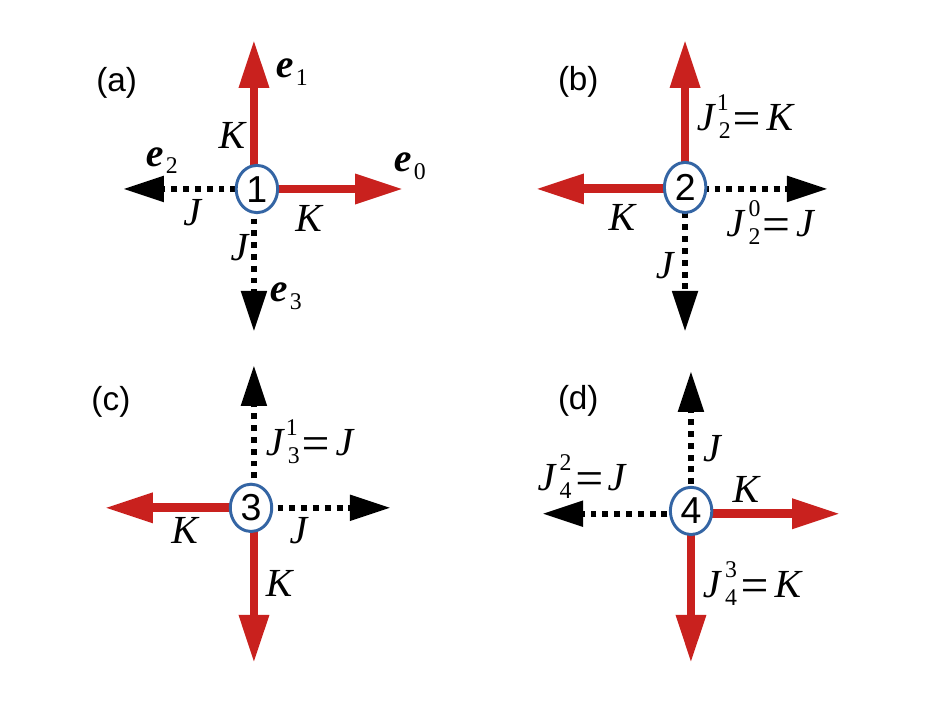}
    \caption{(Color online) Four interaction configurations. (a) $s_{\bf i}=1$:  interaction strength is $K$ in right and up direction (thick line) and  $J$  in  left and down direction (dashed line). (b),(c),(d) corresponds to   $s_{\bf i}=2,3,4.$ 
    In general $s_{\bf i}=s$ interact with its  $k$-th neighbour (at ${\bf i+e}_k$) with strength $J^k_s$ given by Eq. \eqref{eq:Jsk}. For the $q$-state Potts model we consider only first $q$ interaction configurations. 
   }
    \label{fig:spin4}
\end{figure}

{\it The model:}   The model is defined on a   periodic  square lattice  ${\cal L}$ with sites  labeled by ${\bf i} \equiv (x,y)$   with $x,y =1,2,\dots, L.$  The neighbours of  a site ${\bf i}$  is denoted by 
${\bf i} + {\bf e}_k$ where   the unit vector  ${\bf e}_k \equiv (\cos(\frac\pi2k),  \sin(\frac\pi2k)),$   with  $k=0,1,2,3.$
Each site  ${\bf i}$ is occupied by a spin $s_{\bf i}$ which  can be in one of $q$ states, i.e., $s_{\bf i}=1,2,\dots,q.$  A spin   $s_{\bf i}$   interacts with  its  $k$-th neighbors  with 
 interaction strength $J^k_{s_{\bf i}}$ that depends explicitly on both the spin-value $s_{\bf i}$ and   the direction of the neighbour $k.$ (see  Fig.  \ref{fig:spin4} and descriptions there). Energy  of any  configuration  $\{ s_{\bf i} \}$ is given by   
\be
E(\{ s_{\bf i} \}) = -\sum_{ {\bf i}\in {\cal L}}  \sum_{k=0}^3 J^k_{s_{\bf i}} \left(2 \delta_{s_{\bf i}, s_{{\bf i} + {\bf e}_k}} -1\right).  \label{eq:nonrec} 
\ee
Note  that, $J_s^k>0$   represents  ferromagnetic interaction among spins, as energy of the system is lowered  by  making  neighbouring spins  parallel (same).  Further,  we specify the directional interaction  as follows.   
\be
J_s^k= K (\delta_{s,k} + \delta_{s,k+1}) + J (\delta_{s,k-1} + \delta_{s,k-2}). \label{eq:Jsk}
\ee
For example, when  $s_{\bf i}=1$  the interaction occurs along the direction   ${\bf e}_0$ and ${\bf  e}_1$  with strength $K$ (thick lines  in   Fig.  \ref{fig:spin4} (a)) and  the same occurs  with spins in  other  two directions  with strength $J$ (broken lines in Fig. \ref{fig:spin4}(a)). 
The energy can be written  as $E(\{ s_{\bf i} \})= E_x(\{ s_{\bf i} \}) + E_y(\{ s_{\bf i} \})$  where  $E_{x}$ and $E_{y}$ are  the contributions from  bonds oriented  in  $x$  (horizontal) and $y$ (vertical) directions. 
\bea
E_x(\{ s_{\bf i} \}) &=& -\sum_{ {\bf i}\in {\cal L}}   J^x_{s_{\bf i}, s_{{\bf i} + {\bf  e}_0}} 
\left(2 \delta_{s_{\bf i}, s_{{\bf i} + {\bf  e}_0}} -1\right) \cr 
E_y(\{ s_{\bf i} \})  &=&-\sum_{ {\bf i}\in {\cal L}}   J^y_{s_{\bf i}, s_{{\bf i} + {\bf  e}_1}} 
\left(2 \delta_{s_{\bf i}, s_{{\bf i} + {\bf  e}_1}} -1\right). 
\eea
Note that the sum  over the sites  counts  exactly one  horizontal and one vertical bond  with interaction strengths  
\be
J^x_{s,\tilde s} = J^{0}_s +  J^{2}_{\tilde s};~  J^y_{s,\tilde s} = J^{1}_s +  J^{3}_{\tilde s}
\ee
respectively.  The non-reciprocal  nature of  the interaction is  clear  from the fact that  $J^{x,y}_{s,\tilde s} \ne J^{x,y}_{\tilde s,s},$  which is evident from  Eq. \eqref{eq:Jsk}. 

First  we use a dynamics that satisfy detailed balance, where a  configuration  $\{..,s_{\bf i},..\}$ changes $\{ ..,\tilde s_{\bf i},..\}$  with  the  Metropolis rate $r= {\rm Min} (1, e^{-\Delta E})$ where  $\Delta E = E(\{ ..,\tilde s_{\bf i},..\}) - E(\{ ..,s_{\bf i},..\})$.
This dynamics  takes the system to  equilibrium; the partition function   at temperature $T$ ($\beta =\frac{1}{k_B T}$, $k_B$ is  Boltzmann constant) is 
\be 
Q (\beta) =\sum_{\{ s_{\bf i} \}} \prod_{\bf i}   \la s_{\bf i}| M_x  | s_{ {\bf i} +{\bf  e}_0} \ra \prod_{\bf i}   \la s_{\bf i}| M_y  | s_{ {\bf i} +{\bf  e}_1}\ra.
\ee
Here $M_{x,y}$ are $q\times q $   matrix  with elements,
\be
\la s| M_{x,y}  |\tilde s\ra =   e^{\beta J^{x,y}_{s,\tilde s} (2 \delta_{s,\tilde s}-1)}.
\ee
It turns out that  matrices  $M_x$ and $M_y$ do not commute, but they are isospectral; their eigenvalues are,
\bea
\lambda_1=\lambda_2=\dots = \lambda_{q-1}= 2 \sinh(\beta(J+K));\cr  \lambda_q= q\cosh(\beta(J+K)) - (q-2)\sinh(\beta(J+K)).
\eea
It is  easy to check that  both $M_x$ and $M_y$ are isospectral to another  $q\times q$ matrix $M$ with elements  $\la s|M|\tilde s\ra =   e^{\beta \epsilon(2 \delta_{s,\tilde s}-1)},$ that appears in the partition function of the ordinary  $q$-state  Potts model  on a two  dimensional  square lattice with  $H=-\epsilon \sum_{\bf i} \sum_{k=0}^3 (2 \delta_{s_{\bf i}, s_{{\bf i+e}_k}} -1).$
\be
Q_{\rm Potts} (\beta) =\sum_{\{ s_{\bf i} \}} \prod_{\bf i}   \la s_{\bf i}| M  | s_{ {\bf i} +{\bf e}_0} \ra \prod_{\bf i}   \la s_{\bf i}| M | s_{ {\bf i} +{\bf e}_1} \ra.
\ee
Thus, the partition function of the non-reciprocal $q$-state Potts model with equilibrium dynamics can be exactly mapped onto that of the conventional reciprocal Potts model when $\epsilon = K + J$.

\begin{figure}[h]
    \centering
    \includegraphics[width=0.48\linewidth]{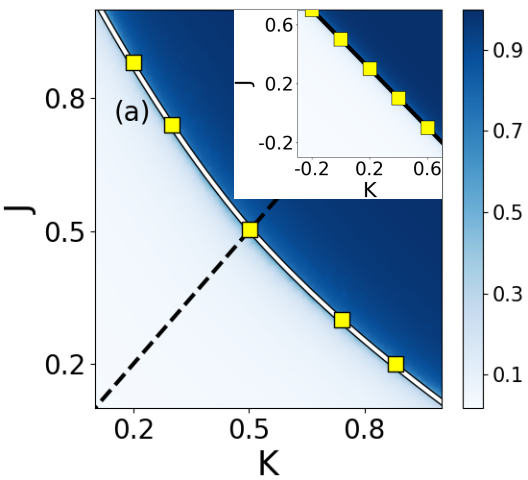}    \includegraphics[width=0.48\linewidth]{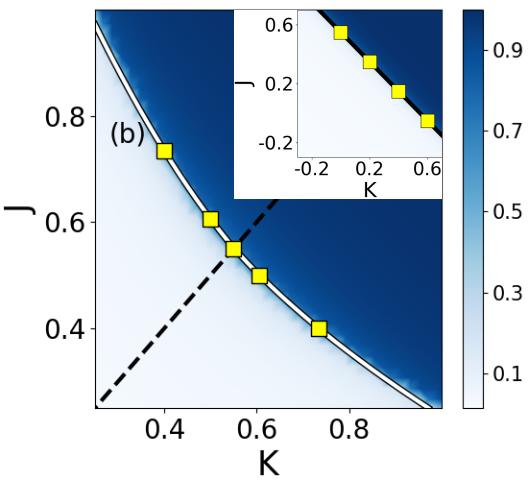}   
    \includegraphics[width=0.48\linewidth]{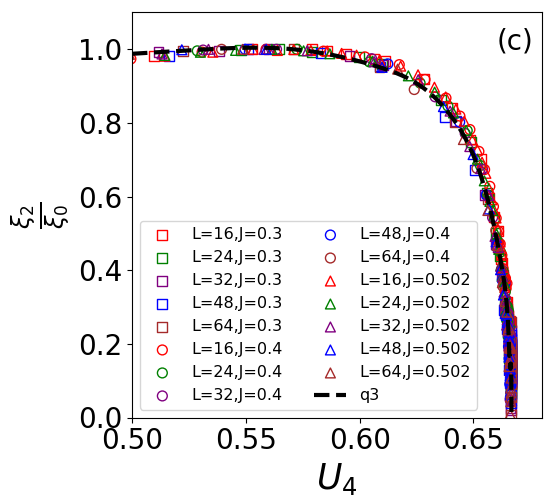}    \includegraphics[width=0.48\linewidth]{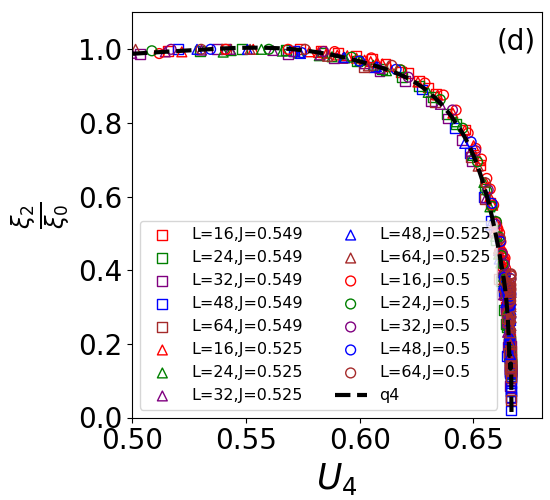}
    \caption{(Color online)(a),(b) Density plots of the order parameter $\phi$ in the $K$-$J$ plane for the non-reciprocal Potts model with selfish non-equilibrium dynamics for $q=3,4$ . The critical line separates the ordered phase (darker colors) from the disordered phase, with symbols marking the critical points at which the critical behavior is analyzed. The resulting critical exponents are listed in Table \ref{table}. The insets show the same data for equilibrium dynamics, where the linear critical line agrees with the prediction from Eq. \eqref{eq:Jc}. (c),(d) Super-universal scaling function: $\xi_2/\xi_0$ vs. the Binder cumulant $U_4$ for $q=3,4$ respectively. The data (symbols), obtained by varying $K$ for different values of $J$ and $L$, collapse onto a single curve, which matches the equilibrium $q=3,4$ Potts model. }
    \label{fig:PPq4}
\end{figure}

The $q$-state Potts models  belong to the $\mathbb{Z}_q$ universality class  with their  critical  points (for a square lattice) and critical exponents  known exactly \cite{Baxter_Book_1982}, 
\bea
\epsilon_c = \ln(1 + {\sqrt q})T; \mu= \frac{2}{\pi}\cos^{-1}\left(\frac{\sqrt q}2\right); \cr
\nu= \frac{2-\mu}{3 -3\mu}; ~ \beta =\frac1{12} (1+\mu); \gamma= \frac{7-4\mu + \mu^2}{6 -6\mu}.
\label{eq:exact}
\eea
We expect  the phase transition   in  the non-reciprocal $q$-state  Potts model  to  occur  at $T=1,$ when interaction   strength $K$ crosses a threshold
\be  K_c = -J +  \ln(1 + {\sqrt q}). \label{eq:Jc}\ee
with  critical exponents for different $q$ given by Eq. \eqref{eq:exact}. 
We verify this from Monte Carlo simulations.  We consider  the usual order parameter of   the Potts model: if number  of sites  having $s_{\bf i}=k$  is $N_k$ we have  $L^2= \sum_{k=1}^q N_k.$  If the largest among  the $\{N_k\}$  is $N_{max},$ the order parameter $\phi$, the susceptibility $\chi,$  and the Binder cumulant $U_4$ are given by 
\bea
&&\phi= \la \mu \ra; ~ \chi=\la \mu^2\ra -  \la \mu\ra^2; U_4= 1-\frac{ \la \mu^4\ra}{3 \la \mu^2\ra^2}; \cr 
&&{\rm where} ~\mu= \frac{1}{q-1} \left(q \frac{ N_{max}}{L^2}-1\right).
\label{eq:phi-chi-U4}
\eea
From  Monte Carlo simulations we obtain $\mu$ and its  moments to construct $U_4, \phi, \chi.$ In the inset of Fig. \ref{fig:PPq4} (a) and (b), we have shown the heat-map of  the order parameter $\phi$ in the $K$-$J$ plane for $q=3,4$, respectively. In each case, the color gradients naturally exhibit a straight line that separates the high $\phi$-values from the low one.  For some of the $J$ values, we also obtain $K_c$ accurately (shown as symbols) from the crossing point of the Binder cumulants $U_4;$ 
they match with Eq.  \eqref{eq:Jc} and fall on  the linear boundary of the order-disorder phase. 
The critical exponents all along this critical line are expected to be  identical to those of the $q$-state Potts model, given  in Eq. \eqref{eq:exact} \cite{Baxter_Book_1982}.

\begin{figure}[h]
    \centering
    \includegraphics[width=0.48\linewidth]{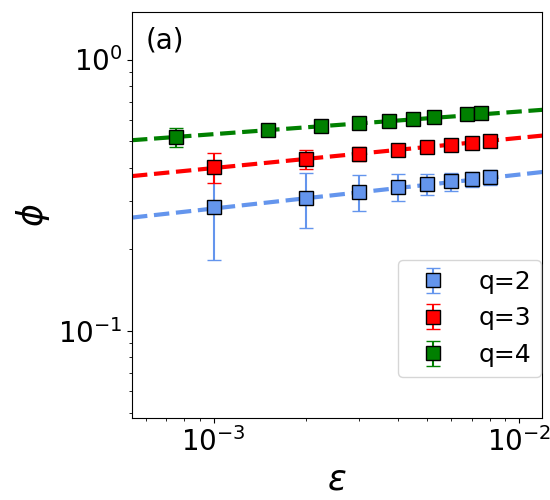} 
    \includegraphics[width=0.48\linewidth]{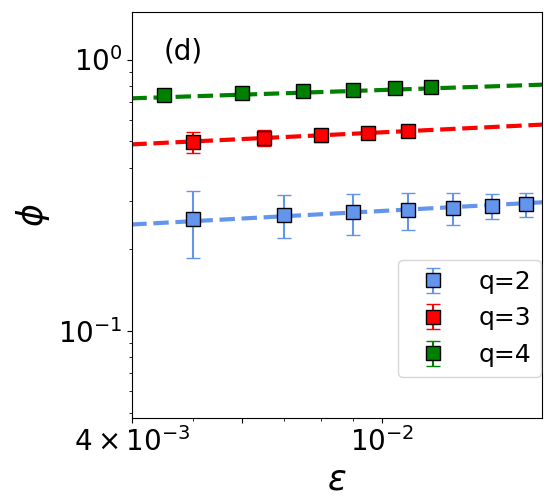}
    \includegraphics[width=0.48\linewidth]{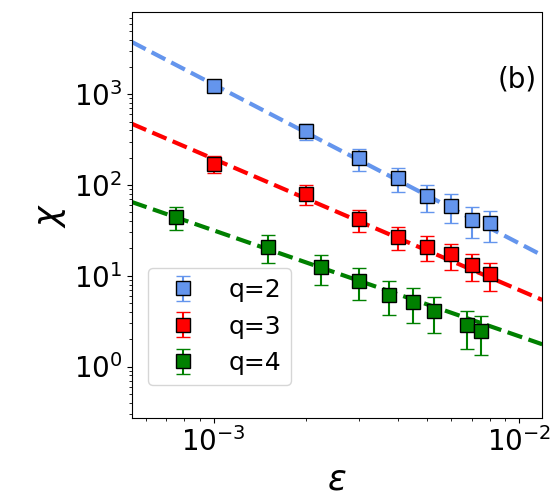}  
    \includegraphics[width=0.48\linewidth]{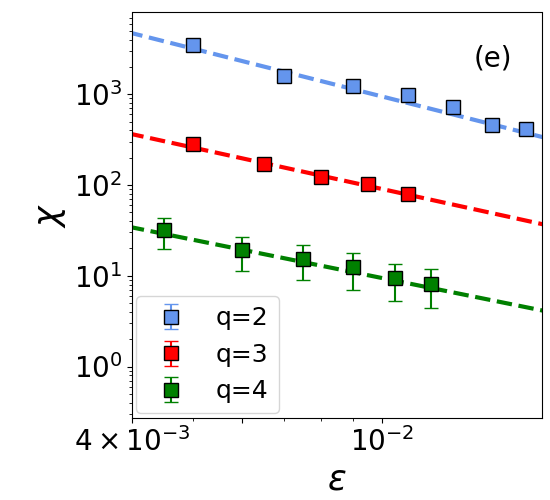}
    \includegraphics[width=0.48\linewidth]{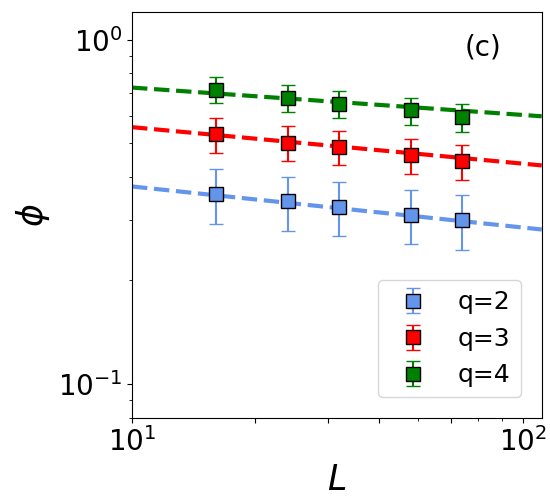}
    \includegraphics[width=0.48\linewidth]{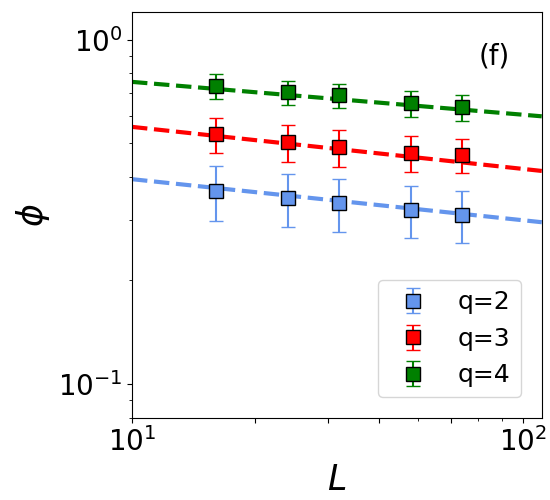}
    \caption{(Color online) (a),(b),(c) Estimate of critical exponents $\beta, \gamma, \beta/\nu$ for non-reciprocal Potts model with equilibrium dynamics, at $J=0.2.$ (a), (b)  shows log-scale plot of $\phi$ vs $\epsilon=K-K_c$  and  $\chi$ vs $\epsilon$ for $L=256.$  (c) $\phi$  vs $L$  at $K=K_c=0.2407 (q=2$), $0.3025 (q=3)$ and $0.3496 (q=4).$   Straight lines of slope $\beta, \gamma, \beta/\nu$   of $\mathbb{Z}_2, \mathbb{Z}_3$ and $\mathbb{Z}_4$  universality class (given in Eq. \eqref{eq:exact}) are drawn along the data for comparison.  (d),(e),(f): The same for  non-equilibrium dynamics  with $J=0.2,0.4,0.5$ respectively for $q=2,3,4,$ resulting in    $K_c=0.762,0.614,0.6065.$ Slope of the best fit lines  provide the  estimate of  $\beta,\nu,\beta/\nu$ (listed in table \ref{table}  for $q=3,4$, and for $q=2,$  we get Ising  exponents $\beta=\frac18=\beta/\nu$ and $\gamma=7/4).$ Data  are averaged over $10^7$ samples.     
    }
    \label{fig:expo}
\end{figure}

We obtain the critical exponents $\beta, \gamma, \frac{\beta}{\nu}$ from finite size scaling \cite{Stanley_Book_1971}  for different critical points (see Supplemental Material \cite{SM}). One  can obtain  it directly from   simulating large system   using the relations: $\phi \sim \epsilon^\beta$, $\chi\sim \epsilon^{-\gamma}$ where $\epsilon= K-K_c$,    and  $\phi\sim L^{\frac\beta\nu}$ at the critical  point $\epsilon=0.$  For $T=1, J=0.2$, log-scale plot of  $\phi$ and $\chi,$  obtained from simulations  $L=256$  for different $\epsilon$  are shown in Fig. \ref{fig:expo} (a),(b); straight lines having slopes $\beta$ and $\gamma$, taken from Eq. \eqref{eq:exact} for different $q$-values, are drawn along the data (symbols) for comparison. A good match suggests that the non-reciprocal interaction does not change the universality of the system when the dynamics follows detailed balance.

We now focus on the non-equilibrium dynamics, commonly referred to as the {\it selfish dynamics} \cite{Avni2024}. In this dynamics, the spins at any site can change their values without affecting the energy of the neighboring spins. Let us rewrite the energy function in Eq. \eqref{eq:nonrec} as
$E(\{ s_{\bf i} \}) = \sum_{ {\bf i}\in {\cal L}} E_{\bf i}(s_{\bf i})$ where
\bea 
E_{\bf i}(s_{\bf i}) = 
- \sum_{k=0}^3 J^k_{s_{\bf i}} \left(2 \delta_{s_{\bf i}, s_{{\bf i} + {\bf  e}_k}} -1\right).  \label{eq:selfE} 
\eea
In the selfish dynamics, a spin $s_{\bf i}$ at ${\bf i}$ changes to a randomly chosen value $\tilde s_{\bf i}\ne s_{\bf i}$ with rate,
\be
r= {\rm Min} (1, e^{-\Delta E_{\bf i}})~ {\rm with}~ \Delta E_{\bf i} = E_{\bf i}(\tilde s_{\bf i}) - E_{\bf i}(s_{\bf i}).
\ee
The term $\Delta E_{\bf i}$ does not include the change of energy felt at neighboring sites  when  $s_{\bf i} \to \tilde s_{\bf i},$ justifying the name -selfish dynamics. This dynamics does not satisfy detailed balance condition wrt energy function \eqref{eq:nonrec} and it evolves the system to a non-equilibrium steady state (NESS).
In NESS, the system exhibits a non-equilibrium phase transition from a disordered to a ordered state characterized by the order parameter $\phi,$ susceptibility $\chi$ and Binder cumulant $U_4$ defined in Eq. \eqref{eq:phi-chi-U4}. For a fixed value of $J,$ the crossing point of $U_4$ vs. $K$ curve for different $L$ gives us the critical value $K_c$ and log-scale plot of $\phi,\chi$ as a function of $\epsilon= K-K_c$ at $L=256$ gives us the estimate of critical exponents $\beta, \gamma.$ In addition, log-scale plot $\phi$ vs $L$ at $K=K_c$ provides $\frac\beta\nu.$ These curves are shown in Fig. \ref{fig:expo} (d),(e) and (f); three different plots in each figure correspond to $q=2,3,4$ and for $J=0.2, 0.4, 0.5$ respectively. For $q=2,$ the estimate of critical exponents match with that of the equilibrium Ising model, whereas for $q=3,4$ they differ from respective exponents of equilibrium Potts model with $q=3,4.$

This led us to investigate the critical behavior all along the critical line shown in Fig. \ref{fig:PPq4} (a),(b) respectively for $q=3,4;$ there, the heat-map of order parameter $\phi$ in $K$-$J$ plane appears in two different colors about the critical line. The resulting critical exponents, obtained from finite size scaling (described in Supplemental Material \cite{SM}) are listed in Table \ref{table}. Clearly, the critical exponents of non-equilibrium non-reciprocal Potts model vary continuously when $q=3$ or $4.$ Note that when $J=K,$ the selfish dynamics is identical to that of the Potts model with equilibrium dynamics and thus, the critical exponents here (3rd row in Table \ref{table}) are given by \eqref{eq:exact}.
\begin{table}[t]

\centering
\begin{ruledtabular}
\begin{tabular}{cc}
$q=3$ & $q=4$ \\ \hline
\begin{tabular}{ccccc}
$J$ & $K_c$ & $\beta$ & $\gamma$ & $\beta/\nu$ \\ \hline
0.300 & 0.738 & 0.160 & 1.586 & 0.168 \\
0.400 & 0.614 & 0.111 & 1.517 & 0.128 \\
0.503 & 0.503 & 0.111 & 1.444 & 0.133 \\
0.614 & 0.400 & 0.111 & 1.517 & 0.128 \\
0.738 & 0.300 & 0.160 & 1.586 & 0.168 \\

\end{tabular} &
\begin{tabular}{ccccc}
$J$ & $K_c$ & $\beta$ & $\gamma$ & $\beta/\nu$ \\ \hline
 0.500 & 0.606 & 0.069 & 1.40 & 0.090 \\
0.525 & 0.576 & 0.080 & 1.44 & 0.100 \\
0.549 & 0.549 & 0.083 & 1.17 & 0.125 \\
0.576 & 0.525 & 0.080 & 1.44 & 0.100 \\
0.606 & 0.500 & 0.069 & 1.40 & 0.090 \\

\end{tabular} \\
\end{tabular}
\end{ruledtabular}
\caption{Table of critical exponents for selfish dynamics.}
\label{table}
\end{table}
Continuously varying exponents often emerge due to the presence of marginal operators. Recent studies \cite{ Delfino2017,Bonati_PRL2019, Indranil_2023, Kanti_2024, Aikya2024} demonstrate that in such cases, the critical behavior forms a superuniversality class associated with the parent fixed point. Specifically, the superuniversal scaling function $U_4 = \mathcal{G}(\xi_2/\xi_0)$ remains invariant along the entire critical line. Here, $\xi_0$ represents the maximum value of the correlation length (finite for any finite $L$), while $\xi_2$ is the second-moment correlation length, defined as
\be
(\xi_2)^2 = \frac{\int_0^\infty r^2 C(r) , dr}{\int_0^\infty C(r) , dr}, \quad \text{where} \quad C(r) = \langle S_{\bf i} \cdot S_{\bf i+r} \rangle - \phi^2. \nonumber
\ee
The spin variables $S_{\bf i}$ are constructed from the Potts state $s_{\bf i} = k$ (with $1 \leq k \leq q$), corresponding to a unit vector representation: $S_{\bf i} = (\cos \frac{2 \pi k}{q}, \sin \frac{2 \pi k}{q})$. Notably, the average magnitude of the total spin ${\bf S}= \frac{1}{L^2}\sum_{\bf i} S_{\bf i}$ in the steady state is given by $\langle \sqrt{ {\bf S \cdot S} }\rangle = \phi$.

To confirm the superuniversality scenario, we compute $\xi_2$ and $\xi_0$ from Monte Carlo simulations for various $K$ values while keeping $J$ and $L$ fixed. The procedure is repeated for different $J,L$ values. Using this data, we construct parametric plots of $\xi_2/\xi_0$ as a function of $U_4$ in Fig. \ref{fig:PPq4}(c) for $q=3$ and  in  Fig. \ref{fig:PPq4}(d) for $q=4$. These plots closely match the reference curve (dashed line) obtained from the standard $q=3$ and $q=4$ equilibrium Potts models with reciprocal interactions.

A few clarifications are in order. For $q=2$, we have assigned the spin configurations (a) and (b) in Fig. \ref{fig:spin4} to $s_{\bf i}=1,2$, respectively. Alternative configurations are equally valid. For instance, one could assign (a) and (c) to $s_{\bf i}=1,2$; under this assignment, the model resembles the vision-cone interactions of Ising spins described in \cite{Garces2024}. In fact, it is straightforward to show that for any of the six (four) possible interaction assignments, the isospectrality argument for equilibrium dynamics remains valid for both $q=2$ and $q=3$ cases. This robustness against interaction assignments reaffirms that non-reciprocity does not alter the universality class of systems with discrete symmetry.

Additionally, interchanging $J$ and $K$ corresponds merely to a different spin assignment, leaving the critical line symmetric. In other words, if $(J, K)$ is a critical point, so is $(K, J)$. This symmetry is evident from Eq. \eqref{eq:Jc} for equilibrium dynamics and is also reflected in Fig. \ref{fig:PPq4} for non-equilibrium dynamics.

In summary, we investigated the effects of non-reciprocal interactions in two-dimensional systems with discrete symmetry, exemplified by the $q$-state Potts model. In equilibrium, the model exhibits continuous phase transitions for $q=2, 3, 4$, breaking discrete symmetries $\mathbb{Z}_2, \mathbb{Z}_3, \mathbb{Z}_4$, respectively, while for $q>4$, the transitions become discontinuous. The universality classes of the continuous transitions are characterized by the critical exponents given in Eq. \eqref{eq:exact}.

When non-reciprocal interactions are introduced, the $q=2$ model (Ising system) undergoes continuous phase transitions belonging to the Ising universality class, irrespective of whether the dynamics satisfy detailed balance (equilibrium) or violate it (selfish dynamics). For $q=3$ and $q=4$, the critical exponents in equilibrium align with the $\mathbb{Z}_3$ and $\mathbb{Z}_4$ universality classes, respectively. However, under selfish dynamics, the system exhibits non-equilibrium phase transitions with continuously varying critical exponents. Despite this variation, a superuniversality class emerges, characterized by invariant super-universal scaling functions along the critical line, even when critical exponents vary.

Overall, our findings reveal that in two dimensions, non-reciprocity does not alter the universality class of phase transitions breaking discrete symmetries. The scenario for systems with continuous symmetries, however, could be fundamentally different. In equilibrium, continuous symmetries cannot be broken to produce true long-range order, yet such systems exhibit exotic phases in two dimensions and exceptional phase transitions in the presence of non-reciprocal interactions \cite{Fruchart2021}.

\bibliography{nonreci.bib}

\end{document}


\title{Supplementary Material for: Non-reciprocal interactions preserve the  universality class of  Potts model}
\author{ Soumya K. Saha and P. K. Mohanty}
\email{pkmohanty@iiserkol.ac.in}
\affiliation {Department of Physical Sciences, Indian Institute of Science Education and Research Kolkata, Mohanpur, 741246 India.}


\date{\today}
\begin{abstract}
In this Supplemental Material, we present a finite size scaling (FSS) analysis \cite{Stanley} to investigate the critical behavior of the non-equilibrium $q$-state Potts model. This includes both the equilibrium dynamics, which obey detailed balance, and the selfish (non-equilibrium) dynamics as defined in the main text.
\end{abstract}
\maketitle

\section{Critical Behavior}
We numerically investigate the critical behavior of the non-reciprocal Potts model defined in the main text using Monte Carlo simulations. Two types of dynamics are employed. The first is an equilibrium dynamics that drives the system to an equilibrium state with a Boltzmann distribution with respect to a non-reciprocal energy function. We explicitly demonstrate that this equilibrium state can also be achieved from a Hamiltonian with reciprocal interactions between spins, where the non-reciprocal interaction parameters $K$ and $J$ uniquely map to the strengths of the reciprocal interactions. This indicates that the non-reciprocal interactions do not alter the critical behavior of the $q$-state Potts model. To verify this explicitly, we calculate the critical exponents through Monte Carlo simulations of the model.

For the $q$-state Potts model \cite{Baxter}, each spin $s_{\bf i}$ can occupy one of $q$ states, $s_{\bf i}=1,2,\dots,q$. In a given configuration, let $N_k$ denote the number of sites with $s_{\bf i}=k$, and let $N_{max}$ represent the largest among the ${N_k}$. The order parameter $\phi$, susceptibility $\chi$, and Binder cumulant $U_4$ of the system are defined as:
\begin{equation}
U_4= 1-\frac{ \la \mu^4\ra}{3 \la \mu^2\ra^2};  ~\phi= \la \mu \ra; ~ \chi=\la \mu^2\ra -  \la \mu\ra^2; ~
{\rm where} ~\mu= \frac{1}{q-1} \left(q \frac{ N_{max}}{L^2}-1\right)
\label{eq:fss}
\end{equation}
 where $L$ is the linear size of the system, and $\langle \dots \rangle$ denotes ensemble averaging. These quantities serve to characterize the phase transitions and critical behavior under both equilibrium and non-equilibrium dynamics.
To obtain the static critical exponents  $\nu,\beta,\gamma$ we   employ  the  finite size scaling analysis \cite{Stanley},
 \begin{eqnarray}\label{eq:scaling_binder}  U_4 = f_{b}(\varepsilon L^{\frac1\nu});~
 \phi = L^{-\frac{\beta}{\nu}}f_\phi(\varepsilon L^{\frac1\nu}); ~ 
 \chi = L^{\frac\gamma\nu}f_\chi(\varepsilon L^{\frac1\nu}).
\end{eqnarray}
  Our approach is as follows: For a fixed value of $J$, we calculate $\phi$, $\chi$, and $U_4$ as defined in Eq. \eqref{eq:fss}. This process is repeated for various system sizes $L = 16, 24, 32, 48, 64$. From the crossing points of the Binder cumulants $U_4$ versus $K$ for different $L$, we determine the critical value $K_c$. Using this, we compute $\epsilon = K - K_c$.

Next, we plot $U_4$ as a function of $\epsilon L^{1/\nu}$ for different $L$ and adjust the fitting parameter ${1/\nu}$ to collapse the curves onto a single master curve. Once $1/\nu$ is determined, we proceed by separately plotting $\phi L^{\beta/\nu}$ and $\chi L^{\gamma/\nu}$ against $\epsilon L^{1/\nu}$. The fitting parameters in these cases are $\beta/\nu$ and $\gamma/\nu$, respectively, which are adjusted to achieve the best data collapse. This method provides accurate estimates of $\frac{1}{\nu}$, $\frac{\beta}{\nu}$, and $\frac{\gamma}{\nu}$.

In all figures below, the inset of panel (a) shows $U_4$ versus $\epsilon = K - K_c$, with the curves crossing at $\epsilon = 0$ (i.e., $K = K_c$). The main panel (a) displays the collapse of $U_4$ as a function of $\epsilon L^{1/\nu}$, yielding an estimate of $\frac{1}{\nu}$. Similarly, panels (b) and (c) show the collapse of $\phi L^{\beta/\nu}$ and $\chi L^{\gamma/\nu}$ versus $\epsilon L^{1/\nu}$, respectively, allowing us to determine $\frac{\beta}{\nu}$ and $\frac{\gamma}{\nu}$. The insets of  panels (b) and (c) shows the  raw-data, before collapse. An independent   estimate  these exponents gives us an opportunity to verify the hyperscaling relation $d\nu= 2\beta+ \gamma$ \cite{Baxter}.

The same methodology is applied to estimate the critical exponents for the selfish non-equilibrium spin dynamics.

\begin{figure}[h]
 \includegraphics[width=.32\textwidth]{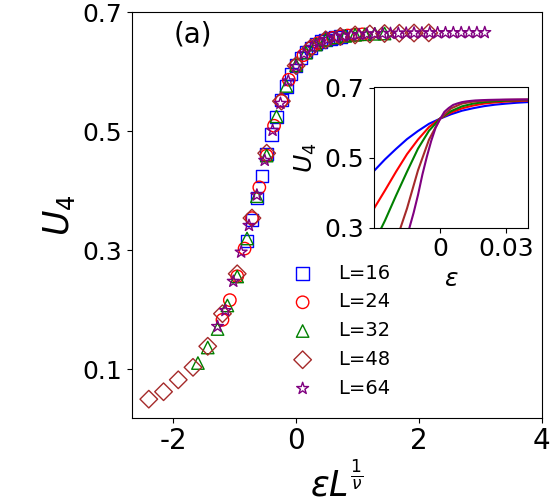}
  \includegraphics[width=.32\textwidth]{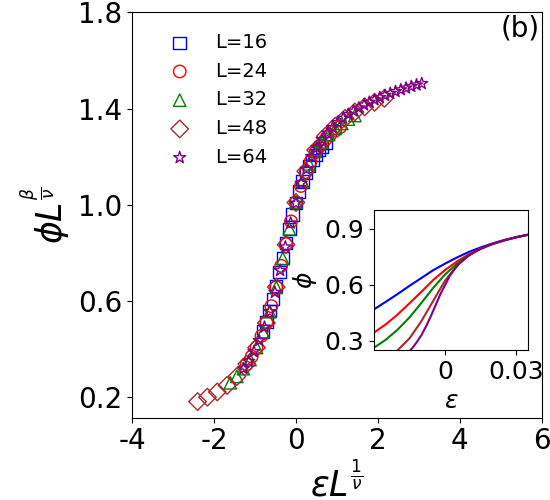}
  \includegraphics[width=.32\textwidth]{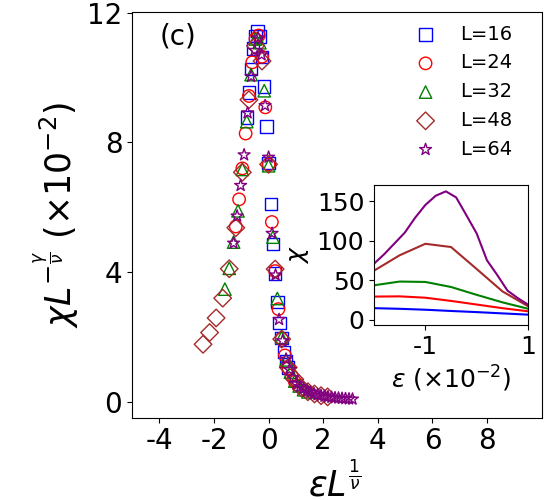}
\caption{From FSS, critical exponents obtained are (a) $\frac{1}{\nu}=1$, (b) $\frac{\beta}{\nu}=0.125$ and (c) $\frac{\gamma}{\nu}=1.75$  for $q=2$ equilibrium dynamics corresponding to the critical point  $J=0.2,K_c=0.2407$}
\end{figure}

\begin{figure}[h]
 \includegraphics[width=.32\textwidth]{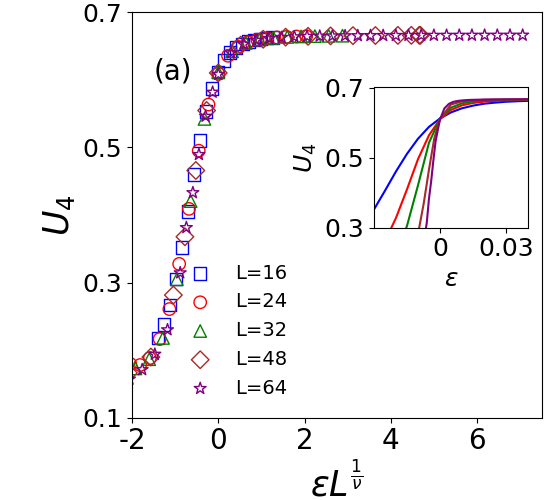}
  \includegraphics[width=.32\textwidth]{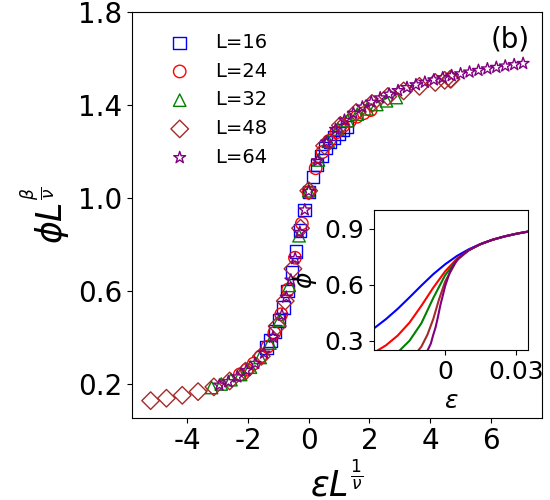}
  \includegraphics[width=.32\textwidth]{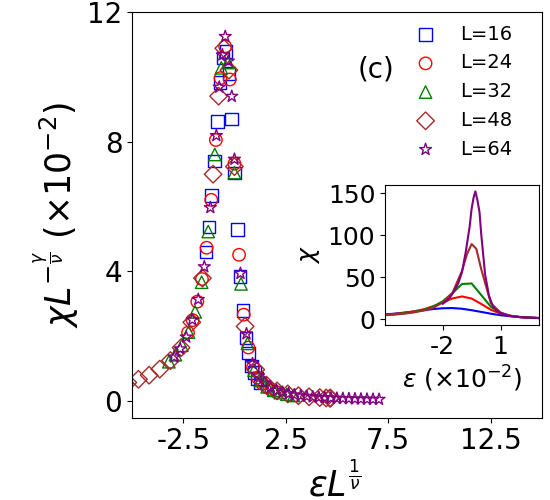}
\caption{From FSS, critical exponents obtained are (a) $\frac{1}{\nu}=1.2$, (b) $\frac{\beta}{\nu}=0.133$ and (c) $\frac{\gamma}{\nu}=1.733$  for $q=3$ equilibrium dynamics corresponding to the critical point  $J=0.2,K_c=0.3025.$}
\end{figure}

\begin{figure}[h]
\includegraphics[width=.32\textwidth]{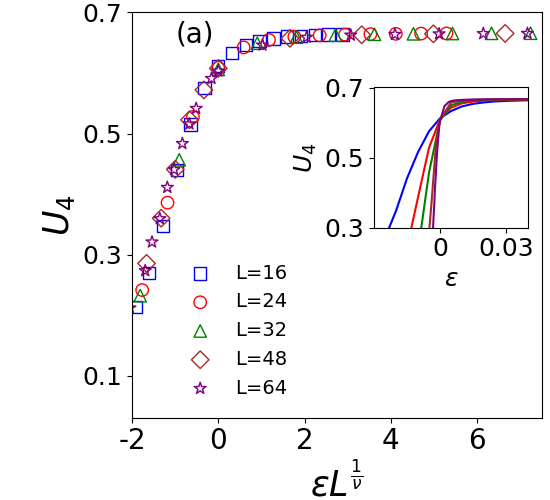}
 \includegraphics[width=.32\textwidth]{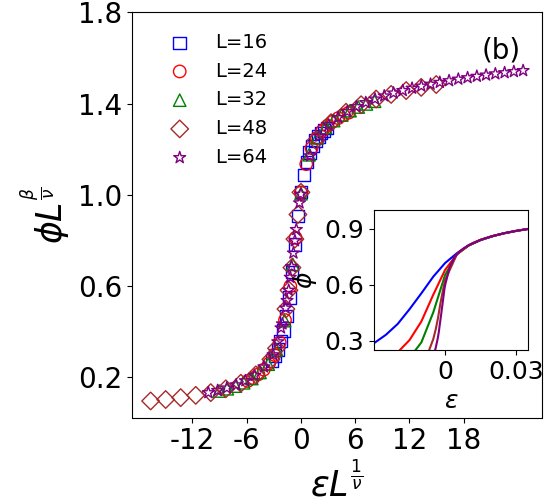}
 \includegraphics[width=.32\textwidth]{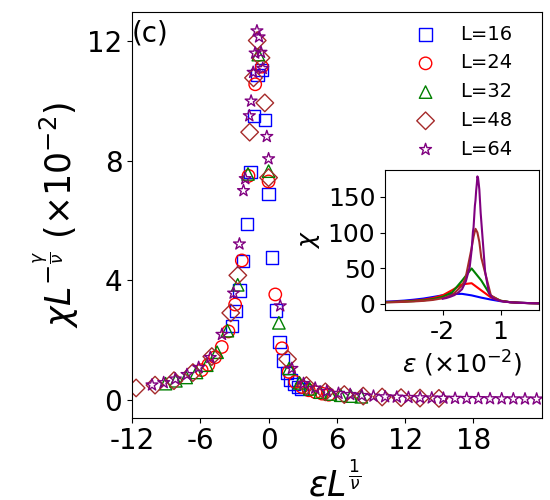}
\caption{From FSS, critical exponents obtained are (a) $\frac{1}{\nu}=1.5$, (b) $\frac{\beta}{\nu}=0.125$ and (c) $\frac{\gamma}{\nu}=1.75$  for $q=4$ equilibrium dynamics corresponding to the critical point  $J=0.2,K_c=0.3493$}
\end{figure}

\begin{figure}[h]
\includegraphics[width=.32 \textwidth]{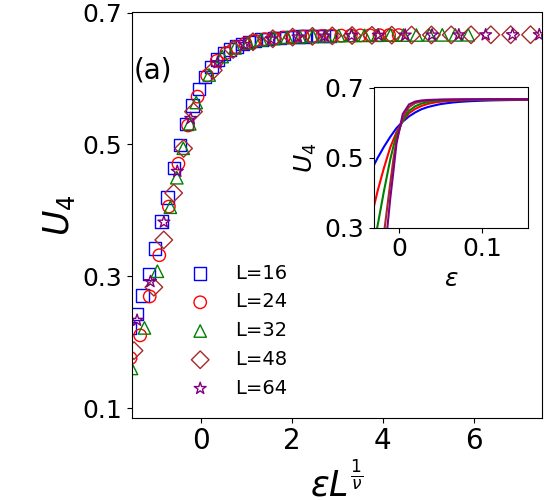}
\includegraphics[width=.32 \textwidth]{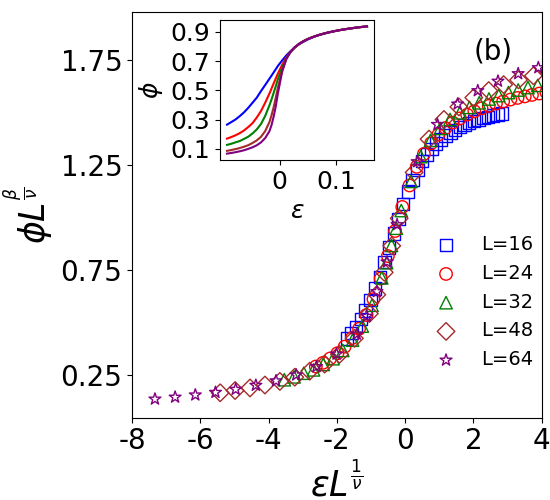}  \includegraphics[width=.32 \textwidth]{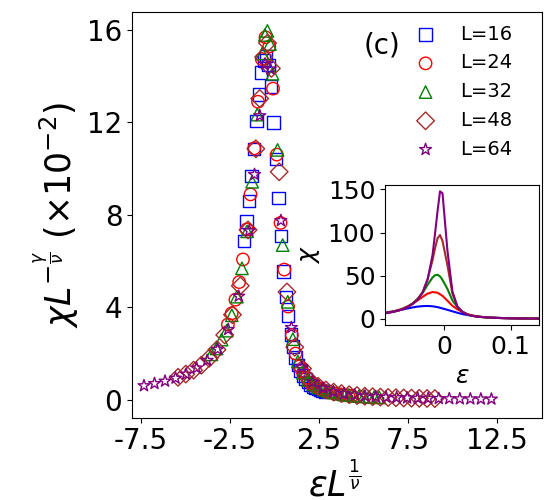}  \caption{From FSS, critical exponents obtained are (a) $\frac{1}{\nu}=1.05$, (b) $\frac{\beta}{\nu}=0.168$ and (c) $\frac{\gamma}{\nu}=1.665$  for $q=3$ selfish dynamics corresponding to the critical point  $J=0.3,K_c=0.738$}
\end{figure}

\begin{figure}[h]
\includegraphics[width=.32\textwidth]{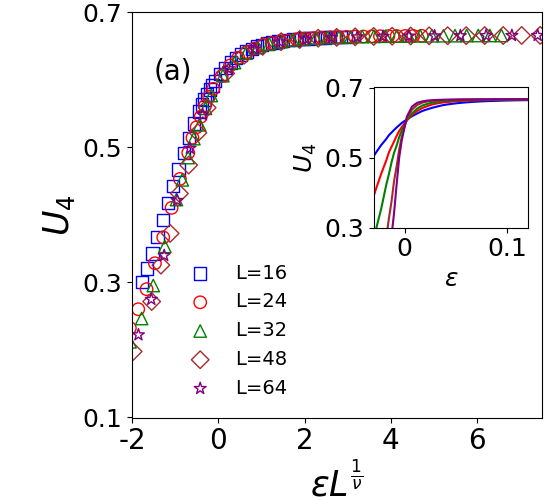}
\includegraphics[width=.32\textwidth]{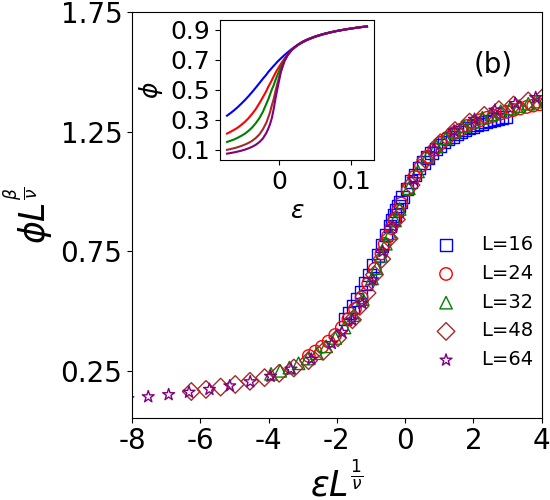}
\includegraphics[width=.32\textwidth]{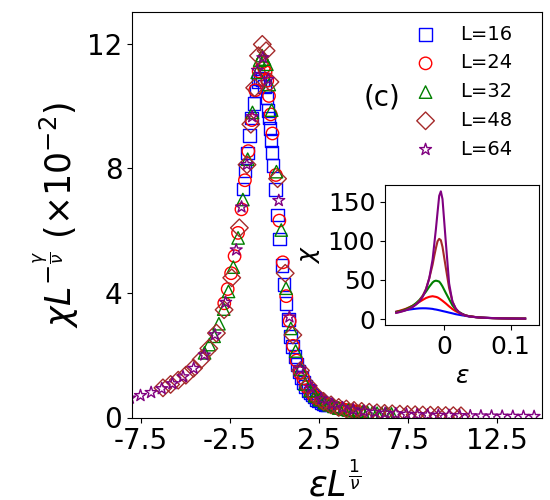}
\caption{From FSS, critical exponents obtained are (a) $\frac{1}{\nu}=1.15$, (b) $\frac{\beta}{\nu}=0.128$ and (c) $\frac{\gamma}{\nu}=1.744$  for $q=3$ selfish dynamics corresponding to the critical point  $J=0.4,K_c=0.614$}  
\end{figure}

\begin{figure}[h]
\includegraphics[width=.32\textwidth]{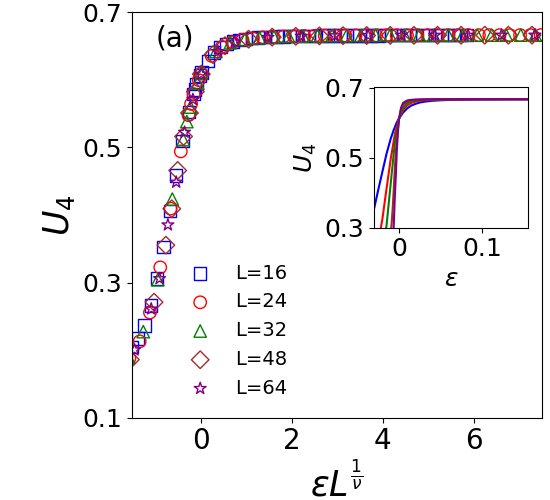}
\includegraphics[width=.32\textwidth]{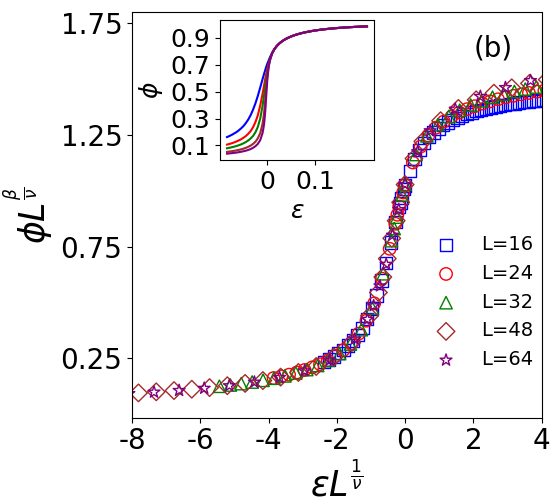}
\includegraphics[width=.32\textwidth]{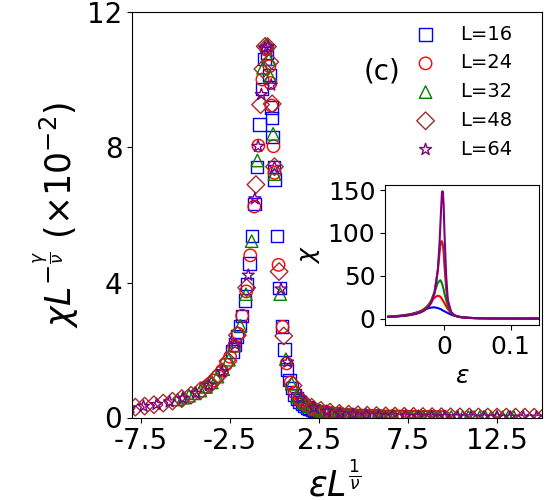}\caption{From FSS, critical exponents obtained are (a) $\frac{1}{\nu}=1.2$, (b) $\frac{\beta}{\nu}=0.133$ and (c) $\frac{\gamma}{\nu}=1.733$  for $q=3$ selfish dynamics corresponding to the critical point  $J=0.5025,K_c=0.5025$}
\end{figure}

\begin{figure}[h]
\includegraphics[width=.32\textwidth]{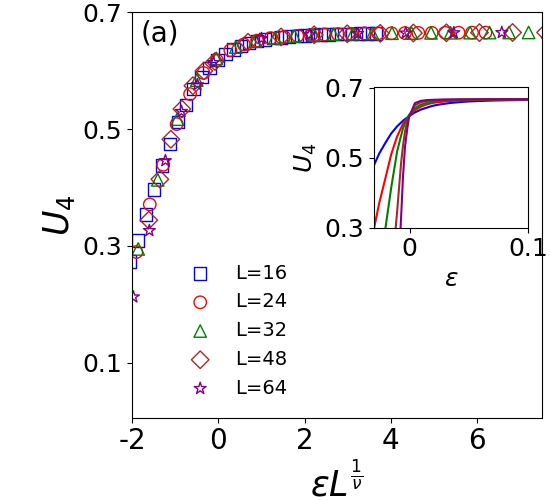}
\includegraphics[width=.32\textwidth]{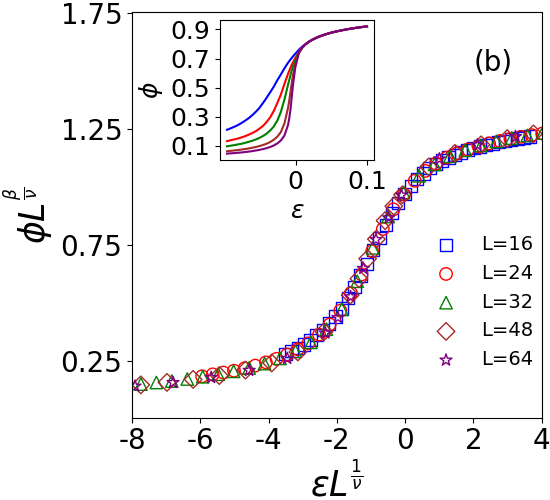}
\includegraphics[width=.32\textwidth]{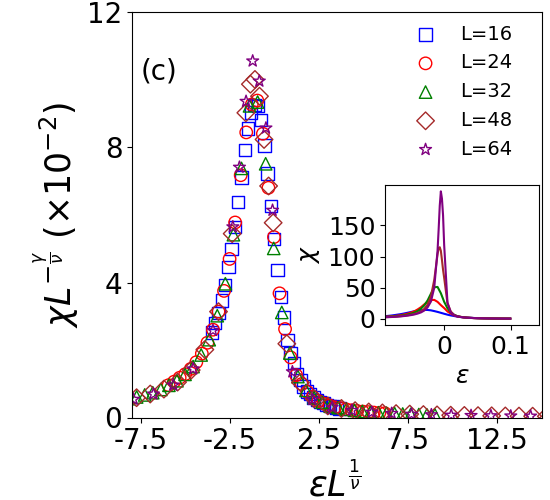}
\caption{From FSS, critical exponents obtained are (a) $\frac{1}{\nu}=1.3$, (b) $\frac{\beta}{\nu}=0.1$ and (c) $\frac{\gamma}{\nu}=1.8$  for $q=4$ selfish dynamics corresponding to the critical point  $J=0.5,K_c=0.6065$}  
\end{figure}

\begin{figure}[h]
   \includegraphics[width=.32\textwidth]{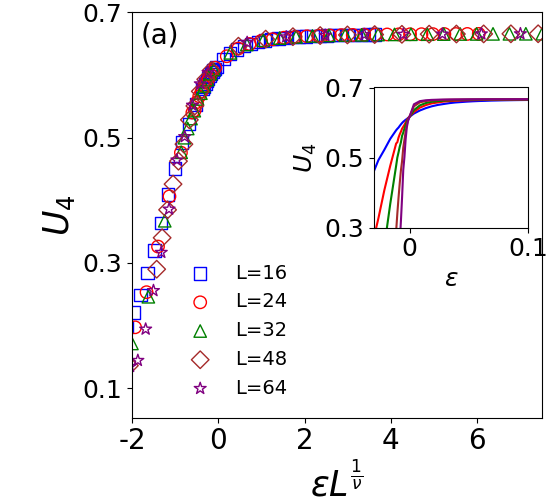}
    \includegraphics[width=.32\textwidth]{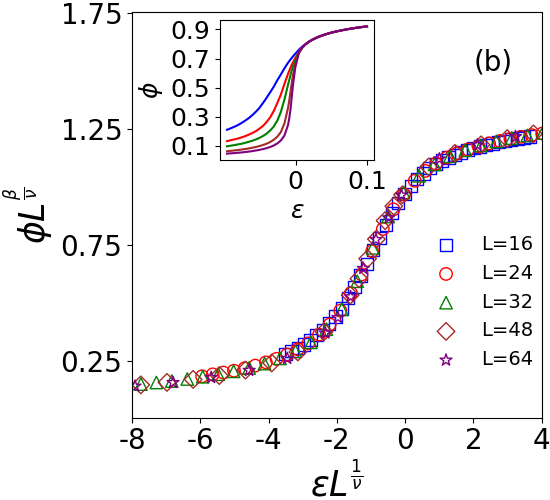}
    \includegraphics[width=.32\textwidth]{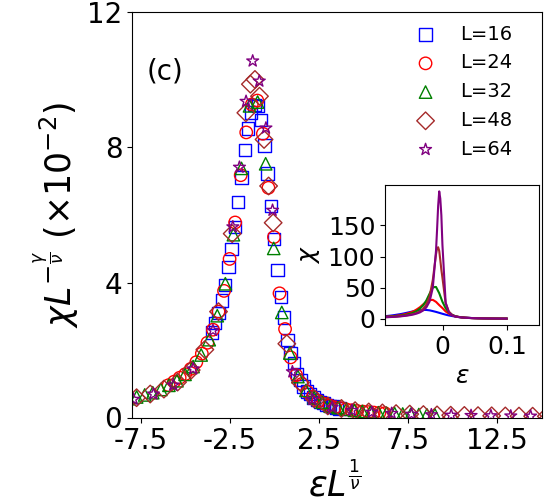}
    \caption{From FSS, critical exponents obtained are (a) $\frac{1}{\nu}=1.25$, (b) $\frac{\beta}{\nu}=0.1$ and (c) $\frac{\gamma}{\nu}=1.8$  for $q=4$ selfish dynamics corresponding to the critical point  $J=0.525,K_c=0.5765$}
\end{figure}

\begin{figure}[h]
\includegraphics[width=.32\textwidth]{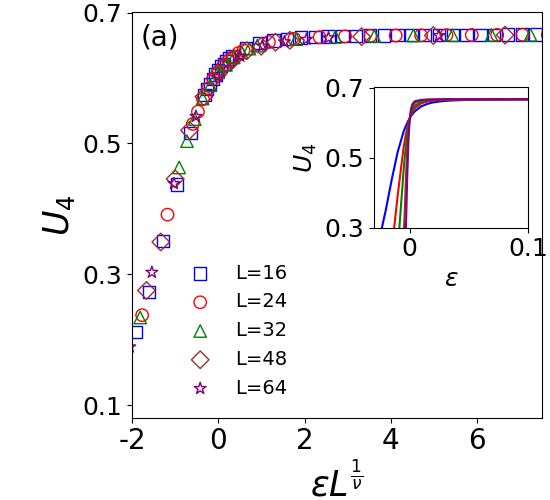}
\includegraphics[width=.32\textwidth]{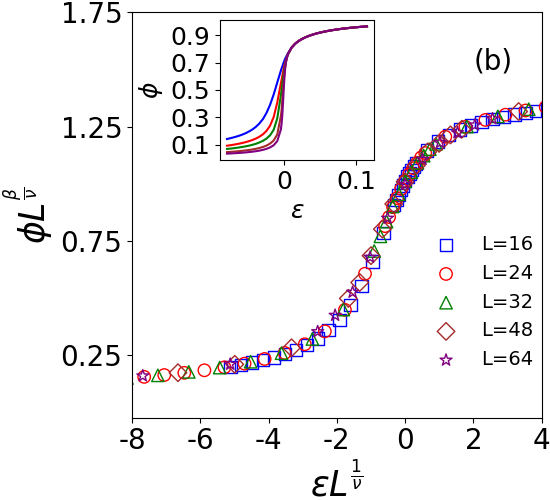}
\includegraphics[width=.32\textwidth]{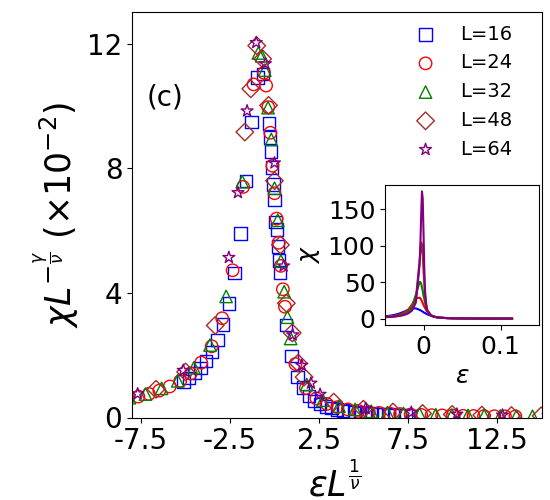}
 \caption{From FSS, critical exponents obtained are (a) $\frac{1}{\nu}=1.5$, (b) $\frac{\beta}{\nu}=0.125$ and (c) $\frac{\gamma}{\nu}=1.75$  for $q=4$ selfish dynamics corresponding to the critical point  $J=0.5495,K_c=0.5495$}
\end{figure}

